\documentclass[5p,times]{elsarticle}
\usepackage{amsmath,amsfonts}
\usepackage{algorithm}
\usepackage{algpseudocode}
\usepackage{array}
\usepackage{textcomp}
\usepackage{stfloats}
\usepackage{url}
\usepackage{verbatim}
\usepackage{graphicx}
\def\BibTeX{{\rm B\kern-.05em{\sc i\kern-.025em b}\kern-.08em
    T\kern-.1667em\lower.7ex\hbox{E}\kern-.125emX}}
\usepackage{balance}
\usepackage{caption}
\usepackage{subcaption}
\usepackage[table,xcdraw,dvipsnames]{xcolor}

\usepackage{amsmath} 
\usepackage{amssymb}  
\usepackage{amsthm}
\theoremstyle{definition}
\newtheorem{definition}{Definition}
\newtheorem{remark}{Remark}
\newtheorem{theorem}{Theorem}

\newcommand{\im}{\textrm{i}}
\newcommand{\R}{\mathbb{R}}

\newcommand{\K}{\mathcal{K}}

\newcommand{\E}{\mathcal{E}}

\newcommand{\tr}{\text{tr}}
\newcommand\brown[1]{{\color{Red}#1}}
\newcommand\green[1]{{\color{OliveGreen}#1}}

\begin{document}

\begin{frontmatter}
\title{Modular Redesign of Mechatronic Systems: Formulation of Module Specifications Guaranteeing System Dynamics Specifications}

\author[aff1]{Lars A.L. Janssen \corref{corresponding}}
\ead{L.A.L.Janssen@tue.nl}

\author[aff1]{Rob H.B. Fey }
\ead{R.H.B.Fey@tue.nl}

\author[aff2]{Bart Besselink }
\ead{B.Besselink@rug.nl}

\author[aff1]{Nathan van de Wouw}
\ead{N.v.d.Wouw@tue.nl}

\cortext[corresponding]{Corresponding author}
\address[aff1]{Dynamics \& Control, Department of Mechanical Engineering,  Eindhoven University of Technology, The Netherlands}
\address[aff2]{Bernoulli Institute for Mathematics, Computer Science and Artificial Intelligence, University of Groningen, The Netherlands}

\begin{abstract}
Complex mechatronic systems are typically composed of interconnected modules, often developed by independent teams. 
This development process challenges the verification of  system specifications before all modules are integrated.
To address this challenge, a modular redesign framework is proposed in this paper.
Herein, first, allowed changes in the dynamics (represented by frequency response functions (FRFs)) of the redesigned system are defined with respect to the original system model, which already satisfies system specifications.
Second, these allowed changes in the overall system dynamics (or system redesign specifications) are automatically translated to dynamics (FRF) specifications on module level that, when satisfied, guarantee overall system dynamics (FRF) specifications. 
This modularity in specification management supports local analysis and verification of module design changes, enabling design teams to work in parallel without the need to iteratively rebuild the system model to check fulfilment of system FRF specifications. 
A modular redesign process results that shortens time-to-market and decreases redesign costs.
The framework's effectiveness is demonstrated through three examples of increasing complexity, highlighting its potential to enable modular mechatronic system (re)design.
\end{abstract}

\begin{keyword}
Modular System (Re)Design \sep
Dynamics, FRF specifications \sep
Systems Engineering \sep
Mechatronics \sep
Incremental Development
\end{keyword}
\end{frontmatter}

\section{Introduction}
\begin{figure}
  	\centering
   	\includegraphics[scale=.7, page=3]{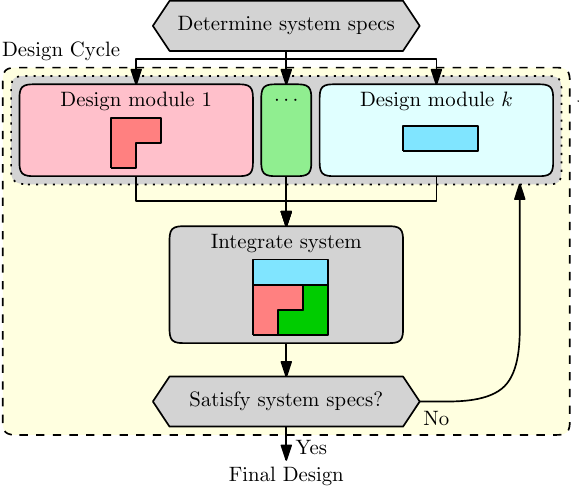}
   	\caption{Integrated design procedure.}
    \label{fig:standard_proc}
\end{figure}

Most complex mechatronic systems consist of multiple interconnected modules/components. 
These systems need to satisfy given design specifications related to their dynamic behaviour on the level of the overall system.
However, in many cases, the modules are developed and designed by independent teams of specialized engineers, which introduces a challenge related to the complexity of the design process.
Namely, if specifications are only provided on a system level, the satisfaction of these specifications can only be verified after all modules have been designed and integrated (or interconnected), see Figure~\ref{fig:standard_proc}.
This may lead to unexpected behavior which could necessitate the need for (expensive) redesigns.
In addition, this leads to a design cycle where delays are unavoidable, as each module design needs to be completed before verification can take place \cite{nielsen2015systems}.

\begin{figure}
  	\centering
   	\includegraphics[scale=.7, page=4]{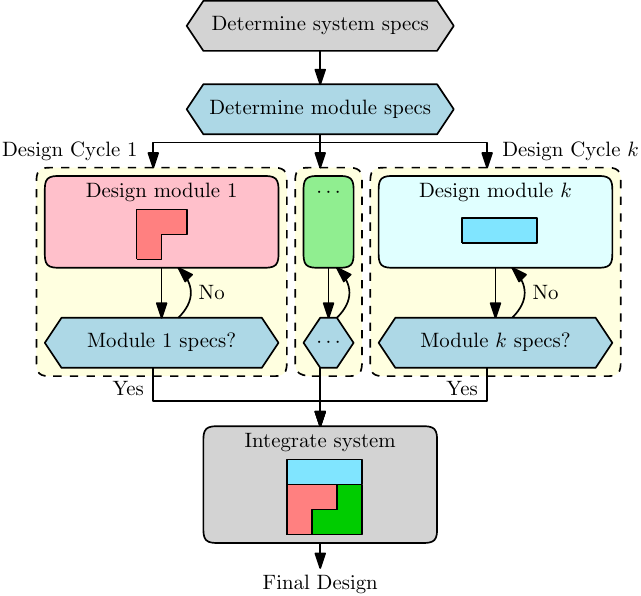}
   	\caption{Modular design procedure.}
    \label{fig:modular_proc}
\end{figure}

Therefore, the aim is often to achieve a modular design approach that decouples the design cycles of modules \cite{baldwin2000design}.
A modular design approach for mechatronic systems requires the formulation of module specifications, given the overall system dynamics specifications \cite{zheng2017multidisciplinary,van2010modular,hamraz2012multidomain}.
Then, when the modules are designed, the satisfaction of their respective specifications can be verified on a module level.
In theory, if all modules are finished and satisfy their specifications, the complete system can directly be integrated and is guaranteed to satisfy system-level specifications (see Figure~\ref{fig:modular_proc}).

Such a modular design approach allows for example parallel design threads of engineering teams, easier replacement of modules of the system and modularity in complexity reduction \cite{giese2004modular,habib2014comparative,barbieri2014model,zheng2019interface}.
However, many challenges persist when aiming to develop a modular design approach.
First, the modules and the interfaces between them need to be clearly defined.
Second, a framework to efficiently determine module specifications based on the overall system specifications is required.
These module specifications need to be feasible and realistic.
Third, if all components satisfy their respective specifications, the satisfaction of the overall system specifications needs to be guaranteed.
Finally, the approach has to be scalable to complex systems with multiple modules with complex interactions.

In general, all of these challenge make ``first time right'' design often infeasible in practice. 
In addition, system specifications could change during (or even after) the design process which could require updates in the design.
Therefore, the need arises for a modular \emph{re}design approach in which existing designs are already available but one or more components of the system need to be replaced/changed.
Such an approach is also called continuous, evolutionary or incremental development \cite{grimheden2013can,de2021process,maier1998architecting}.

In a modular redesign procedure, system-level specifications need to be translated to module-level specifications such that proposed updated module designs can be verified without the need to test the updated system as a whole.
In \cite{janssen2023modular} and \cite{janssen2022modular}, a modular approach is introduced for another purpose, i.e., modular model complexity management.

In \cite{janssen2022modeselect}, this approach was introduced specifically for the complexity management of interconnected
structural dynamics models.
In these works, requirements on the accuracy of a desired reduced-order interconnected model are translated to accuracy requirements on reduced-order subsystem models.
In turn, satisfaction of these FRF requirements for each subsystem guarantees the required accuracy of the reduced-order interconnected system model.
Specifically, requirements are given as bounds on the maximum allowed changes in the frequency response functions (FRFs).
Such bounds are also used in control systems to guarantee for example the stability or performance of a system in the presence of uncertainties, see e.g., \cite{iwasaki2005time,iwasaki2007feedback} or \cite{nordebo1999semi}.
As such, methods from the field of robust control \cite{zhou1998essentials} have been proven useful to support modular model complexity management. 

The main contribution of this work is to show how this line of reasoning can also be exploited to develop a novel, fully modular redesign framework which guarantees the satisfaction of specifications on the system-level input-output behavior, described by a FRF.
This is achieved by first introducing a general modular modelling framework in which FRF representations of the modules' dynamics are interconnected to obtain the overall system dynamics (in terms of FRFs).
In addition, frequency-based system-level specifications are introduced, which define how much the dynamics of the interconnected system are allowed to change in the frequency domain.
With these specifications at hand, we show that module specifications of a similar nature can be automatically computed that, when satisfied, guarantee satisfaction of the overall system specifications.
These module specifications enable a modular redesign process where proposed design changes in the modules can be analysed and verified locally (i.e., on module level), which enables a work-flow in which design teams can work fully in parallel.
This may significantly reduce time-to-market and decrease redesign costs.

The proposed redesign framework is demonstrated on three mechatronic use cases of increasing complexity: 1) an illustrative academic example in the form of a two-mass-spring-damper system, 2) a six component pillar-and-plate benchmark model and 3) the model of an industrial wire bonder used in the semi-conductor industry.

The remainder of this paper is organized as follows. 
Section~\ref{sec:framework} introduces the modular modelling framework. 
In Section~\ref{sec:system_specs}, a way to represent redesign specifications in the frequency domain using FRFs is introduced.
In Section~\ref{sec:module_specs}, the main contribution of this paper will be discussed: How to translate system FRF specifications to module FRF specifications that guarantee the required system-level dynamic behaviour. 
We will illustrate this approach on three different use case systems in Section~\ref{sec:examples} and close with conclusions in Section~\ref{sec:conclusion}.

\section{General framework for modular modelling}
\label{sec:framework}
To enable model-based modular redesign of systems, we will introduce a general modelling framework that can be used to model interconnected systems. 
Here, we focus on the dynamic behaviour of such systems having linear dynamics and model the system as a set of interconnected FRFs. 

In this framework, we assume that each of the $k$ original modules can be modelled as a multiple-input multiple-output (MIMO) FRF given by 
\begin{align}
y^{(j)}(i\omega) = G^{(j)}(i\omega)u^{(j)}(i\omega)
\end{align}
for $j=1,2,\dots,k$ and where $\omega \in \R$ represents frequency. 
Modular redesign would generally lead to a change in FRF, which is represented as
\begin{align}
\hat{y}^{(j)}(i\omega) = \hat{G}^{(j)}(i\omega)\hat{u}^{(j)}(i\omega)
\end{align}
for $j=1,2,\dots,k$ for all $\omega \in \R$. 
We assume that both the original and redesigned modules have the same number of $m_j$ inputs and $p_j$ outputs.
Therefore, the complex matrix $G^{(j)}(i\omega)$ and $\hat{G}^{(j)}(i\omega)$ have the same dimensions ($p_j \times m_j$).

\begin{figure}
  	\centering
   	\includegraphics[scale=.7, page=1]{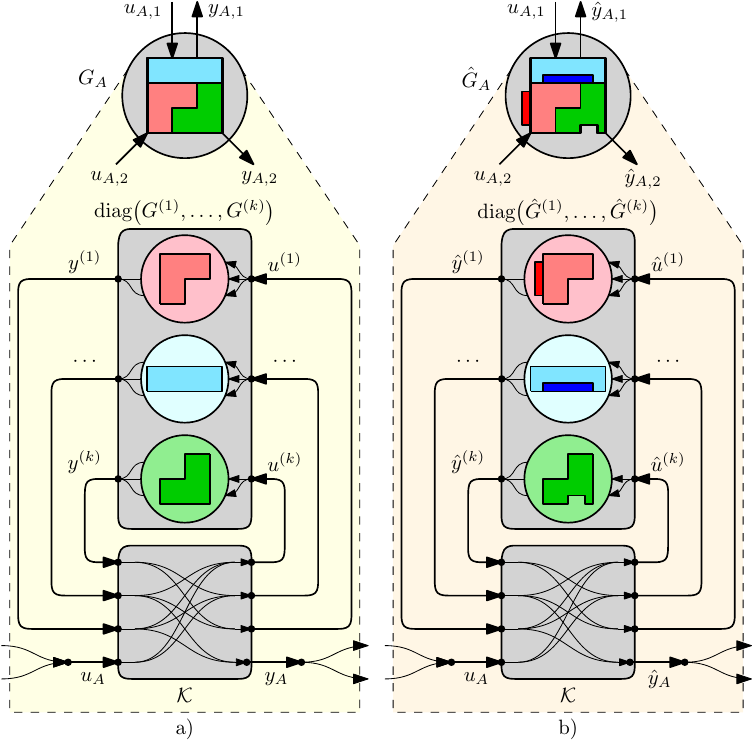}
   	\caption{A schematic example of the modular modelling framework with a fixed interconnected structure $\K$ where a) given the original system model and b) the redesigned system model.}
    \label{fig:modular_model}
\end{figure}

To enable a systematic approach for connecting the modules, the $k$ module FRFs are collected in a block diagonal FRF,  
\begin{align}
\label{e:Gb}
G_B(i\omega) &= \textrm{diag}\big(G^{(1)}(i\omega),\dots,G^{(k)}(i\omega)\big)\\
\hat{G}_B(i\omega) &= \textrm{diag}\big(\hat{G}^{(1)}(i\omega),\dots,\hat{G}^{(k)}(i\omega)\big)
\end{align}
for the original modules and redesigned modules, respectively, for all $\omega \in \R$. 
Then, the interconnection between modules is modelled as signals from outputs of modules ($y_B := [y^{{(1)}^\top},\dots,y^{{(k)}^\top}]^\top$ or $\hat{y}_B(i\omega) := [\hat{y}^{{(1)}^\top},\dots,\hat{y}^{{(k)}^\top}]^\top$) to inputs of (other) modules ($u_B(i\omega) := [u^{{(1)}^\top},\dots,u^{{(k)}^\top}]^\top$ or $\hat{u}_B(i\omega) := [\hat{u}^{{(1)}^\top},\dots,\hat{u}^{{(k)}^\top}]^\top$).
This is captured in the interface matrix $\K_{BB}$, as is common with interconnected LTI systems in the field of control \cite{sandberg2009model,reis2008survey}.
Furthermore, we define the system's $m_A$ external input signals by $u_A(i\omega)$ and its $p_A$ external output signals by $y_A(i\omega)$.
These are connected to the modules via matrices $\K_{AB}$ and $\K_{BA}$, respectively.
The complete interconnection structure $\K$ is then given by
\begin{align}
\label{eq:sigmac}
\left[\begin{array}{c}
u_B \\ y_A 
\end{array}\right] &= \K \left[\begin{array}{c}
y_B \\ u_A 
\end{array}\right] \text{ and } 
\left[\begin{array}{c}
\hat{u}_B \\ \hat{y}_A 
\end{array}\right] = \K \left[\begin{array}{c}
\hat{y}_B \\ u_A 
\end{array}\right] \\ \text{ with }
\K &= \left[\begin{array}{cc}
\K_{BB} & \K_{BA} \\
\K_{AB} & 0
\end{array}\right]. \nonumber
\end{align}
Note that we assume here that the interconnection structure $\K$ is the same for the original and the redesigned modules.
Using this modular modelling framework illustrated in Figure~\ref{fig:modular_model}, we can define the system's original FRF from $u_A$ to $y_A$ as
\begin{align}
\label{eq:Gc}
G_A(i\omega) &:= \K_{AB}G_B(i\omega)\left( I - \K_{BB}G_B(i\omega)\right)^{-1}\K_{BA}, 
\end{align}
and the redesigned FRF from $u_A$ to $\hat{y}_A$
\begin{align}
\label{eq:Gchat}
\hat{G}_A(i\omega) &:= \K_{AB}\hat{G}_B(i\omega)\left( I - \K_{BB}\hat{G}_B(i\omega)\right)^{-1}\K_{BA}.
\end{align}
\begin{remark}
\label{rem:rigid}
Typically, in the field of structural dynamics, rigid interfaces between physical components are used \cite{craig2000coupling,de2008general}.
However, with rigid interfaces, in contrast to the flexible interfaces used in this work, algebraic constraint equations are required to model the interconnected system.
The framework that will be introduced in this paper currently does not allow for such algebraic constraint equations.
As a solution, a very stiff coupling can be used to model almost rigid interconnections between modules.
\end{remark}

\section{User-defined system FRF specifications}
\label{sec:system_specs}
\begin{figure}
  	\centering
   	\includegraphics[trim={0.8cm .1cm 1.1cm .6cm},clip,scale=.7]{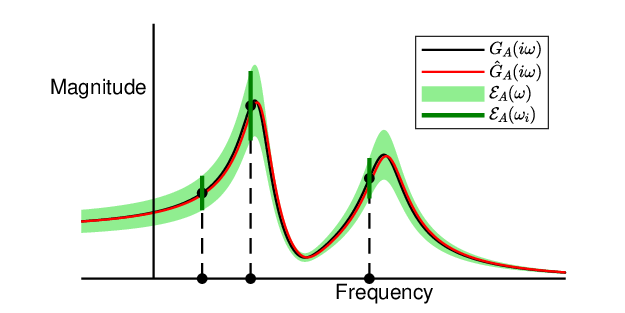}
   	\includegraphics[trim={2.6cm 2cm 3.5cm .8cm},clip,scale=.7]{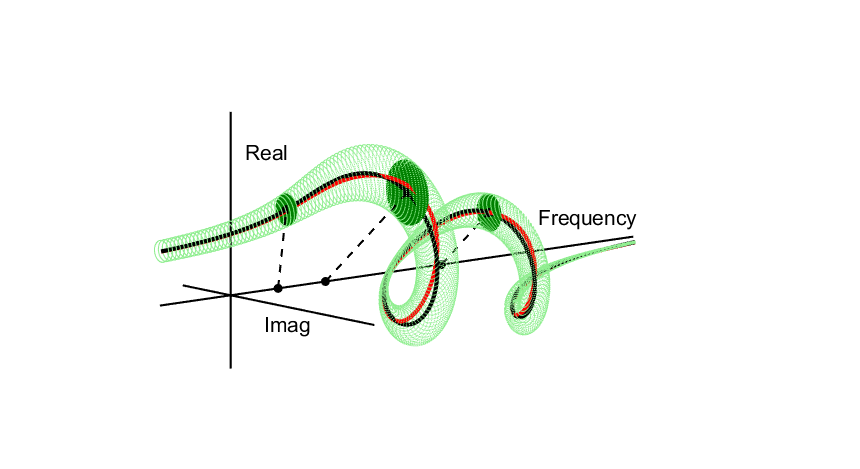}
   	\caption{Two equivalent visualizations of an original system FRF $G_A(i\omega)$, specifications $\E_A(\omega)$, and a redesigned system FRF $\hat{G}_A \in \E_A$. In addition, three specific frequency points are highlighted to show that $\E_A(\omega)$ not only restricts the magnitude, but also the phase of $\hat{G}_A(i\omega)$.}
    \label{fig:req_illu}
\end{figure}

To enable the exploitation of the approach introduced in \cite{janssen2023modular}, we define FRF specifications on the dynamics of the system in the frequency domain.
For the purpose of modular redesign, we pose the following definition for design specifications on the dynamic behaviour in the frequency domain:
\begin{definition}
\label{def:system_spec}
For FRF system specifications $\E_A(\omega)$ given by
\begin{align}
\label{eq:Ec_bound}
\E_A(\omega) & := \Big\{ \hat{G}_A(i\omega) \ \big|  \nonumber \\
 &\|V_A(\omega)(G_A(i\omega)-\hat{G}_A(i\omega))W_A(\omega)\| < 1 \Big\},
\end{align}
the redesigned modular system is considered to satisfy its specifications, denoted by $\hat{G}_A \in \E_A$, if and only if its FRF $\hat{G}_A(i\omega) \in \E_A(\omega)$ for all $\omega \in \Omega$.
Here, $G_A(i\omega)$ is the original system FRF, and $V_A(\omega)$ and $W_A(\omega)$ are positive, diagonal, frequency-dependent scaling matrices, and $\omega$ is evaluated at the frequencies of interest, defined by $\Omega \subseteq \R$.
\end{definition}
\begin{remark}
In this paper, we denote the Euclidean norm or 2-norm of a matrix $A$ as $\|A\|$ which is equal to its largest singular value $\bar{\sigma}(A)$. 
If we denote $\|G(i\omega)\|$, we imply that we simply obtain the 2-norm of $G(i\omega)$ at each frequency $\omega$ which, for SISO systems, is identical to the magnitude of $G(i\omega)$.
\end{remark}
Definition~\ref{def:system_spec} imposes specifications on the system-level that are effectively a bound on the allowed change of the original FRF $G_A(i\omega)$ to the redesigned FRF $\hat{G}_A(i\omega)$ at specific frequency points.
Here, the scaling matrices $V_A(\omega)$ and $W_A(\omega)$ are used to scale the allowed change in the dynamics of the system for each input-to-output pair at each $\omega \in \Omega$.
This allows for a fully flexible redesign specification.
To illustrate how these specifications restrict the allowed change in the dynamics from $G_A(i\omega)$ to $\hat{G}_A(i\omega)$, two equivalent representations of a specification on a SISO example FRF are shown in Figure~\ref{fig:req_illu}. 
Note that for MIMO systems, Definition~\ref{def:system_spec} still applies, but its meaning is difficult to visualize as (\ref{eq:Ec_bound}) restricts all input-output pairs at the same time.

Note that the overall system specification $\E_A(\omega)$ is user-defined, i.e., the system engineer can select the frequency points $\Omega$ and the corresponding weighting matrices $W_A(\omega)$ and $V_A(\omega)$. 
Practically, this means that the system engineer selects the frequency points and the radius of the circles visualized in Figure~\ref{fig:req_illu} at these frequencies. 
For example, a system engineer might consider specifications on the interconnected system that are a tight bound on the FRF at low frequencies and at frequencies that are prominent in the reference trajectories of the system in operation, while at other (higher) frequencies, a less restrictive specification is allowed.

\section{Computation of module specifications with assembly specification guarantees}
\label{sec:module_specs}
In this section, we will show 1) how the system FRF specifications $\E_A$ defined in Definition~\ref{def:system_spec} can be used to automatically generate module FRF specifications $\E^{(j)}$ and 2) that if these specifications are satisfied for all modules, i.e., $\hat{G}^{(j)} \in \E^{(j)}$, then $\hat{G}_A \in \E_A$ is guaranteed.
First, similar to the system specifications, we define the module specifications.
\begin{definition}
\label{def:module_spec}
For FRF module specifications $\E^{(j)}(\omega)$ given by
\begin{align}
\label{eq:Ej_bound}
\E^{(j)}(\omega) &:= \Big\{ \hat{G}^{(j)}(\im\omega) \ \big| \nonumber \\
 &\| (W^{(j)}(\omega))^{-1}(\hat{G}^{(j)}-G^{(j)})(\im\omega)(V^{(j)}(\omega))^{-1} \| \leq 1 \Big\},
\end{align}
any redesigned module $j = 1,2,\dots,k$ is considered to satisfy its FRF specifications, denoted by $\hat{G}^{(j)} \in \E^{(j)}$, if and only if its FRF $\hat{G}^{(j)}(i\omega) \in \E^{(j)}(\omega)$ for all $\omega \in \Omega$. 
Here, $G^{(j)}(i\omega)$ is the FRF of the original module $j$, and $V^{(j)}(\omega)$ and $W^{(j)}(\omega)$ are positive, diagonal, frequency-dependent scaling matrices, and $\omega$ is evaluated at the frequencies of interest, defined by $\Omega \subseteq \R$.
\end{definition}
Note that the module specifications in Definition~\ref{def:module_spec} have the same characteristics as the system specification in Definition~\ref{def:system_spec}.
Namely, they allow for a maximum change in dynamics in the FRF $\hat{G}^{(j)}(\omega)$ with respect to $G^{(j)}(\omega)$ based on the values in the matrices $V^{(j)}(\omega)$ and $W^{(j)}(\omega)$.

To enable a modular redesign approach, we aim to relate the system specifications $\E_A$ to the module specifcations $\E^{(j)}$.
To achieve this, we define the nominal system
\begin{align}
\label{eq:N}
N(i\omega)&:=\left[\begin{array}{cc}
\K_{BB}(I-G_B(i\omega)\K_{BB})^{-1} & (I-\K_{BB}G_B(i\omega))^{-1} \K_{BA}\\ 
\K_{AB}(I-G_B(i\omega)\K_{BB})^{-1}&O
\end{array}\right],
\end{align}
which comprises only of the original modules and interconnection structure.
Furthermore, based on the system and module specifications (\ref{eq:Ec_bound}) and (\ref{eq:Ej_bound}), respectively, we define
\begin{align}
\label{eq:V}
V(\omega) &= \textrm{diag}\big(V^{(1)}(\omega),\dots,V^{(k)}(\omega),V_A(\omega)\big), \text{and} \\
\label{eq:W}
W(\omega) &= \textrm{diag}\big(W^{(1)}(\omega),\dots,W^{(k)}(\omega),W_A(\omega)\big). 
\end{align}
Then, we can pose the following theorem that relates the satisfaction of module specifications to the satisfaction of the assembly specification.
\begin{theorem}
\label{the:top_down}
Consider the nominal system in (\ref{eq:N}) and the specification-related scaling matrices in (\ref{eq:W}).
If there exists 
\begin{align}
\label{eq:D}
(D_\ell, D_r) \in \mathbf{D} 
\end{align}
such that
\begin{align}
\label{eq:LMI}
\left[\begin{array}{cc}
W^{-2}(\omega)D_r^{-1} 	& 	N^H(\im\omega) \\
N(\im\omega)				& 	V^{-2}(\omega)D_\ell
\end{array}\right] &\succ 0
\end{align}
holds\footnote{$A^H$ denotes the conjugate transpose of $A$.}, where, given that $I_n$ denotes an identity matrix of dimension $\R^{n\times n}$,
\begin{align}
\mathbf{D} &:= \Big\{ (D_\ell, D_r) \ \Big| \ d_1,\dots,d_k, d_A \in \R_{>0}, \\ \nonumber
& \qquad \qquad D_\ell = \textrm{diag}\left( d_1 I_{p_1},\dots,d_{k} I_{p_k}, d_AI_{m_A} \right), \\ \nonumber
& \qquad \qquad D_r = \textrm{diag}\left( d_1 I_{m_1},\dots,d_{k} I_{m_k}, d_AI_{p_A} \right)\Big\},
\end{align}
then, satisfaction of the module specifications for all $\omega \in \Omega$, i.e., 
\begin{align}
\hat{G}^{(j)} \in \E^{(j)}
\end{align}
for all $j=1,2,\dots,k$, implies satisfaction of the system specifications for all $\omega \in \Omega$, i.e., 
\begin{align}
\hat{G}_A \in \E_A.
\end{align}
\end{theorem}
\begin{proof}
We can define 
\begin{align}
\label{eq:Ej}
E^{(j)}(i\omega) &= \hat{G}^{(j)}(i\omega)-G^{(j)}(i\omega), \textrm{ and} \\
\label{eq:EA}
E_A(i\omega) &= \hat{G}_A(i\omega)-G_A(i\omega).
\end{align}
Therefore, by substituting (\ref{eq:Ej}) in (\ref{eq:Ej_bound}) and (\ref{eq:EA}) in (\ref{eq:Ec_bound}), Theorem~\ref{the:top_down} becomes equivalent to the constraints of \cite[Theorem 3.1]{janssen2022modeselect} which in turns follows from \cite[Theorem 3.4]{janssen2022modular}.
The proofs in \cite{janssen2022modular} in turn follow from definitions on the upper bound on the structured singular value \cite{packard1993complex}, a mathematical tool used in the field of robust control.
\end{proof}
Theorem~\ref{the:top_down} gives a mathematical approach to verify whether module properties imply system properties, which is the core idea of this approach.
However, note that the matrix inequality (\ref{eq:N}) is not linear in the parameters, $V$, $W$, $D_\ell$ and $D_r$ which makes it difficult to optimize over these parameters. 
We will propose an alternating optimization algorithm that alternates between solving two different linear matrix inequalities (LMIs) to obtain a solution.
Specifically, Theorem~\ref{the:top_down} is used to optimize over the module weights $V^{(j)}$ and $W^{(j)}$ to allow for module redesigns with redesign spaces that are as large as possible.
Therefore, in the following Algorithm~\ref{alg:top_down}, we show how Theorem~\ref{the:top_down} can be used to automatically generate a set of FRF-based specifications $\E^{(j)}(i\omega)$ for all modules $j=1,2,\dots,k$ for any $\omega\in\Omega$ on the basis of the assembly specification $\E_A(i\omega)$.
An example of such a specification is given in Remark~\ref{rem:gamma}. 

\begin{algorithm}
\caption{Computation of Module Specification}\label{alg:top_down}
\textbf{Input:} The original module model FRFs $G^{(j)}(i\omega)$ for all $j=1,2,\dots,k$ and interconnection structure $\K$ to obtain $N$ as in (\ref{eq:N}), and user-defined system specifications $\E_A(\omega)$ as in Definition~\ref{def:system_spec} in terms of $V_A(\omega)$ and $W_A(\omega)$ for a discrete set of frequencies of interest $\omega \in \Omega$. 
\begin{algorithmic}
\State Initialize $D_r^\star := I$, $D_\ell^\star := I$, $i := 0$, and define a stopping criterion $\epsilon \ll 1$.
\For{all $\omega \in \Omega$}
\Repeat
\State $i := i + 1$
\State \parbox[t]{210pt}{Fix $D_r^\star$, $D_\ell^\star$, $V_A(\omega)$ and $W_A(\omega)$, solve the following semi-definite optimization problem:
\begin{align}
\label{eq:beta}
\begin{array}{ll}
\min\limits_{\brown{\beta_i}\in\R} & \brown{\beta_i} := \tr\left(\brown{V^{-2}(\omega)}\right) + \tr\left(\brown{W^{-2}(\omega)}\right),\\
\text{s.t.} & \left[\begin{array}{cc}
\brown{W^{\text{-}2}(\omega)}{D_r^\star}^{\text{-}1} & N^H(\im\omega) \\
N(\im\omega)				& 	\brown{V^{\text{-}2}(\omega)}D_\ell^\star
\end{array}\right] \succ 0.\\
\end{array}
\end{align}
Note that here, $V$ and $W$ are as given in (\ref{eq:V}) and (\ref{eq:W}), respectively, and, as $V_A(\omega)$ and $W_A(\omega)$ are given, only $V^{(j)}(\omega)$ and $W^{(j)}(\omega)$ are optimization parameters.
Define $V^\star(\omega)$, $W^\star(\omega)$ and $\beta_i^\star$ as the solution to the red optimization parameters in (\ref{eq:beta}).\strut}
\State \parbox[t]{210pt}{Fix $V^\star(\omega)$, $W^\star(\omega)$, solve the following semi-definite optimization problem:
\begin{align}
\label{eq:schur}
\begin{array}{ll}
\min\limits_{\green{\delta_i}\in\R} &  \begin{array}{rl}N(i\omega){(W^\star(\omega))}^{2}\green{D_r}N^H(i\omega) & \\ -{(V^\star(\omega))}^{-2}\green{D_\ell} &\prec \green{\delta_i},
\end{array}\\
\text{s.t.} & (\green{D_\ell}, \green{D_r}) \in \mathbf{D}.
\end{array}
\end{align}
Define $D_{\ell}^\star, D^\star_r$ and $\delta_i^\star$ as the solution to the green optimization parameters in (\ref{eq:schur}), which in the next iteration allows for a further reduction of $\beta_i$.\strut}
\Until{{$\beta_{i-1}^\star - \beta_i^\star < \epsilon$}}
\EndFor
\end{algorithmic}
\textbf{Output:} Module specifications $\E^{(j)}(\omega)$ for all $\omega \in \Omega$ as in Definition~\ref{def:module_spec} for all $j=1,2,\dots,k$ given the obtained $V^{(j)}(\omega)$ and $W^{(j)}(\omega)$ from (\ref{eq:beta}) for which it holds that if $\hat{G}^{(j)} \in \E^{(j)}$ for all $j=1,2,\dots,k$, then $\hat{G}_A \in \E_A$ is guaranteed.
\end{algorithm}

In Algorithm~\ref{alg:top_down}, (\ref{eq:beta}) is designed with the aim of automatically reducing allowed changes in individual elements within the scaling matrices $V(\omega)$ and $W(\omega)$ that play a crucial role in the interconnected system which, as a result, gives more allowed change in elements that are less crucial for the satisfaction of the specification of the interconnected system. Furthermore, note that (\ref{eq:schur}) is simply the Schur complement \cite{zhang2006schur} of (\ref{eq:N}) where, additionally, the right-hand-side of the matrix inequality, which was $0$, is substituted by $\delta_i$.

\begin{remark}
\label{rem:gamma}
As an example of a system specification $\E_A$ we consider in this paper a frequency-dependent maximum on the relative difference between ${G}_A(i\omega)$ and $\hat{G}_A(i\omega)$, given by $\gamma(\omega)$, i.e., 
\begin{align}
\label{eq:ex_req}
\frac{\|{G}_A(i\omega) - \hat{G}_A(i\omega)\|}{\|{G}_A(i\omega)\|} < \gamma(\omega),
\end{align}
for all $\omega \in \Omega$. 
Then, we take $W_A(\omega)$ and $V_A(\omega)$ such that (\ref{eq:ex_req}) holds for all $\omega \in \Omega$.
For a SISO system this could for example be $V_A(\omega) = W_A(\omega) = \sqrt{\gamma(\omega)\|{G}_A(i\omega)\|}$.
More examples of system specifications are given in~\ref{app:ex}.
\end{remark}

\begin{remark}
\label{rem:cost_function}
Note that the objective function (\ref{eq:beta}) employed in Algorithm~\ref{alg:top_down} can be modified to obtain a different distribution of the specifications of the modules.
For example, in \cite{janssen2023modular}, it is suggested that the cost function can be easily expanded to 
\begin{align}
\label{eq:improved_beta}
\bar{\beta} = \sum_{j=1}^k \alpha_j\left(\tr(V^{(j)}(\omega))^{-2} + \tr(W^{(j)}(\omega))^{-2}\right),
\end{align}
where the variable $\alpha_j$ is introduced as a weight, providing us with a means to control how the specifications for the modules are distributed within the solution in Algorithm~\ref{alg:top_down}.
\end{remark}

\begin{figure}
  	\centering
   	\includegraphics[scale=.7, page=7]{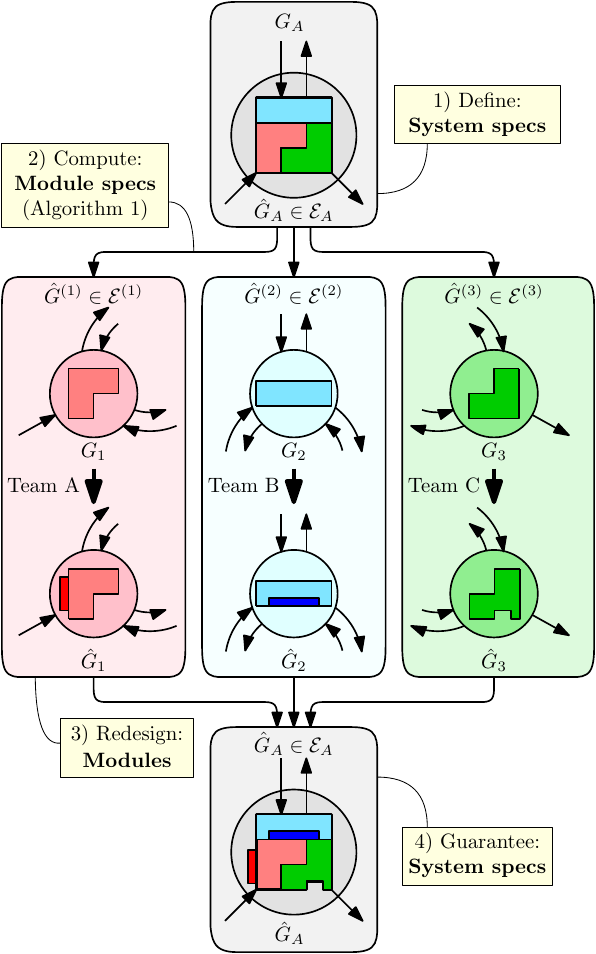}
   	\caption{Visualization of the proposed work-flow to enable modular redesign with guarantees on the system specifications.}
    \label{fig:modular_redesign}
\end{figure}

Note that Algorithm~\ref{alg:top_down} is a formalization of the top-down approach proposed in \cite{janssen2023modular}, which was focussed on model complexity reduction and not system redesign.
This algorithm enables the system engineer to automatically formulate module FRF specifications that, when satisfied, guarantee the satisfaction of the given system FRF specifications. 
Summarizing, this results in the following work-flow:
\begin{enumerate}
\item \textbf{Define system specifications}: The system engineer defines system FRF specifications $\E_A(\omega)$ for a selection of frequencies $\omega \in \Omega$ as in Definition~\ref{def:system_spec}. Practically, this means that $V_A(\omega)$ and $W_A(\omega)$ are defined such that a bound around the FRF of the original system is created that may not be violated, as illustrated in Figure~\ref{fig:req_illu}. This bound should reflect the requirements on the redesigned system.
\item \textbf{Compute module specifications}: Given $G_A(i\omega)$ and $\E_A(\omega)$, Algorithm~\ref{alg:top_down} can be used to automatically generate module FRF specifications $\E^{(j)}(\omega)$ as in Definition~\ref{def:module_spec} for all modules $j=1,2,\dots,k$. 
\item \textbf{Redesign modules}: Given $G^{(j)}(i\omega)$ and $\E^{(j)}(\omega)$, specialists (in engineering teams responsible for a module) can propose redesigns of their respective module. 
These redesigned modules only need to satisfy their local FRF specifications, i.e., the redesigned FRF $\hat{G}^{(j)}(i\omega)$ should satisfy $\hat{G}^{(j)}(i\omega) \in \E^{(j)}(\omega)$ for all $\omega\in\Omega$. Therefore, the design teams can work fully in parallel.
\item \textbf{Redesigned system}: Following from Theorem~\ref{the:top_down}, if $\hat{G}^{(j)} \in \E^{(j)}$ for all $j=1,2,\dots,k$, then $\hat{G}_A \in \E_A$. Therefore, if any redesigned module satisfying its local FRF specifications is integrated into the (redesigned) system, satisfaction of the overall system FRF specifications is guaranteed.
\end{enumerate}
This proposed work-flow is visualized in Figure~\ref{fig:modular_redesign}.

In the introduction of this paper we stated several key requirements to enable a modular redesign approach.
So far, we have shown that the proposed approach introduces a modular modelling framework, user-defined system specifications, and synthesis of module FRF specifications that can guarantee system FRF specifications. 
However, to show that this approach also has an acceptable level of conservativeness and is scalable to systems on industrial scale , we will demonstrate the proposed modular redesign approach on three use cases of increasing complexity.
In addition, we will use these use cases to highlight the advantages of this approach.

\section{Use case systems}
\label{sec:examples}
To show the potential of the proposed modular redesign approach, we will demonstrate the method on three use cases:
\begin{enumerate}
\item \textbf{Two DOF mass-spring-damper system:} In this academic example, we illustrate how FRF specifications are translated from system-level to module-level.
Furthermore, we discuss a potential cause of conservativeness in the approach and propose a solution for this.
\item \textbf{Plate pillar model:} In the Matlab-implemented benchmark system `\verb+"platePillarModel+' \cite{pillarplatemodel}, we show the computational advantage of the modular approach and how relative importance of modules can be analyzed.
\item \textbf{Wirebonder system:} In a use case concerning the motion stages of an industrial wirebonder system, we show how the approach can be implemented on an industrial-scale model.
\end{enumerate}

\subsection{Two DOF mass-spring-damper model}
\begin{figure}
  	\centering
   	\includegraphics[scale=.7, page=5]{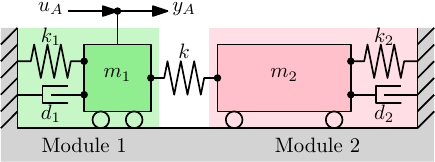}
   	\caption{Schematic representation of the two DOF mass-spring-damper use case system.}
    \label{fig:two_mass}
\end{figure}

\begin{figure}
  	\centering
   	\includegraphics[scale=.7]{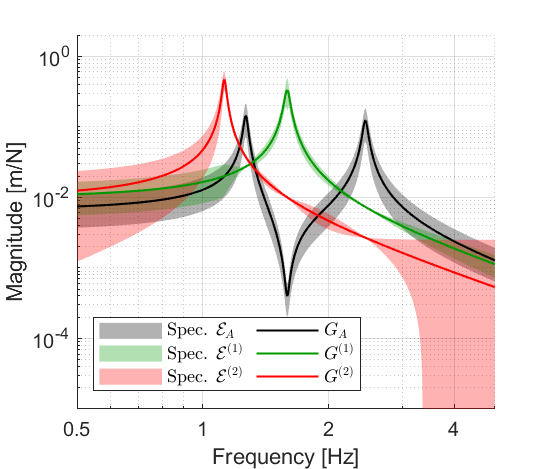}
   	\caption{Magnitudes of the FRFs of the two DOF mass-spring-damper system $|G_A(i\omega)|$, the module FRFs $|G^{(1)}(i\omega)|$ and $|G^{(2)}(i\omega)|$, the user-defined system specifications $\E_A(\omega)$ and the module specifications $\E^{(1)}(\omega)$ and $\E^{(2)}(\omega)$ obtained using Algorithm~\ref{alg:top_down} for $\gamma(\omega) = 0.5$ for all $\omega \in \Omega$.}
    \label{fig:two_mass_FRF}
\end{figure}

The two DOF mass-spring-damper model  illustrated in Figure~\ref{fig:two_mass} is a simple two-module system that we will use to illustrate some of the properties of the modular redesign Algorithm~\ref{alg:top_down}.
To apply this algorithm, we need to supply the FRFs $G^{(j)}(i\omega)$ of the modules, the interconnection structure $\K$, and define system specifications $\E_A(\omega)$.
The module FRFs from module input force to module output position of the two modules are given by 
\begin{align}
G^{(j)}(i\omega) = \frac{1}{-m_j\omega^2 + d_ji\omega + k_j}
\end{align}
for $j = 1,2$.
Note that the two module input forces $u^{(1)}$ and $u^{(2)}$, and the two model outputs $y^{(1)}$ and $y^{(2)}$, used to model the spring-interconnection between the masses, are columnwise stored in $u_B$ and $y_B$, respectively.
The system's external input force $u_A = u^{(1)}$ [N] acts on the inertia in module 1 and the system's external output position is given by $y_A = y^{(1)}$ [m], i.e., we have a SISO system. 
Therefore, the interconnection structure matrix defined in (\ref{eq:sigmac}) is given by
\begin{align}
\K = \left[\begin{array}{cc|c}
-k & k & 1\\
k & -k & 0\\
\hline
1 & 0 & 0
\end{array}\right]
\end{align}
where $k$ is the stiffness of the spring forming the interconnection between the modules.
The overall system FRF $G_A(i\omega)$ from $u_A$ to $y_A$ is then given as in (\ref{eq:Gc}).
Furthermore, we select the following parameter values: $k_1 = k_2 = 100$ N/m, $d_1 = d_2 = 0.3$ Ns/m, $m_1 = 1$ kg, $m_2 = 2$ kg and $k = 90$ N/m.
Finally, we define that any redesigned system FRF $\hat{G}_A(i\omega)$ needs to satisfy the FRF specification $\hat{G}_A(i\omega) \in \E_A(\omega)$ according to Definition~\ref{def:system_spec}.
Specifically, as explained in Remark~\ref{rem:gamma}, we consider a specification as in (\ref{eq:ex_req}) given by two different constant values ($0.05$ and $0.5$) for $\gamma(\omega)$ for all $\omega \in \Omega$, where $\Omega$ is a set of 1000 logarithmically-spaced frequency points between $0.5$ Hz and $5$ Hz. 
Therefore, to obtain system specifications $\E_A$ as in Definition~\ref{def:system_spec}, $W_A(\omega) = V_A(\omega) = \sqrt{\gamma(\omega)\|{G}_A(i\omega)\|}$ for all $\omega \in \Omega$.
Algorithm~\ref{alg:top_down} is then used to automatically obtain module specifications $\E^{(j)}(\omega)$ as in Definition~\ref{def:module_spec} for $j=1,2$.
The resulting system and module FRFs for $\gamma(\omega)=0.05$ for all $\omega\in\Omega$ and their respective FRF specifications are presented in Figure~\ref{fig:two_mass_FRF}. 

To test how the distribution of specifications on a module level affects the allowed design space of each of the modules, we will apply modular redesign by replacing $G^{(1)}(i\omega)$ and $G^{(2)}(i\omega)$ by $\hat{G}^{(1)}(i\omega)$ and $\hat{G}^{(2)}(i\omega)$, respectively.
Specifically, we will investigate the allowed change in masses $m_1$ and $m_2$ in the modules.

With the proposed work-flow given in the previous section, we allow any redesigned module $\hat{G}^{(j)}(i\omega)$ that satisfies $\hat{G}^{(j)} \in \E^{(j)}$.
Therefore, design teams can work independently on the redesign of their module. 
However, being able to guarantee a priori that module changes guarantee the satisfaction of the system FRF specifications $\mathcal{E}_A$, may come with the trade-off that the module FRF specifications $\mathcal{E}^{(j)}$ can be conservative. 
To explore the amount of conservativeness introduced by the proposed method in this simple example system, we will aim to find the allowed masses $m_1$ and $m_2$ that satisfy the FRF specification $\E_A$ using two approaches:
\begin{itemize}
\item \textbf{Modular approach:} This approach follows the proposed work-flow of the previous section, i.e., by finding the values for $m_j$ that satisfy the module-level specifications $\hat{G}^{(j)} \in \E^{(j)}$ for $j = 1,2$. To find the full domain of $m_1$ and $m_2$ that satisfy the $\hat{G}_A \in \E_A$, $\bar{\beta}$ is used to find all distributions of $\E^{(j)}$ that satisfy $\E_A$, as suggested in (\ref{eq:improved_beta}) in Remark~\ref{rem:cost_function}.
\item \textbf{Brute-force approach:} All combinations of masses $m_1$ and $m_2$ are evaluated on the integrated system specification $\hat{G}_A \in \E_A$. Note that this requires evaluation of the system dynamics as a whole.
\end{itemize}
\begin{figure}
	\centering	
   	\includegraphics[scale=.7,trim={0 0.7cm 0 1cm},clip]{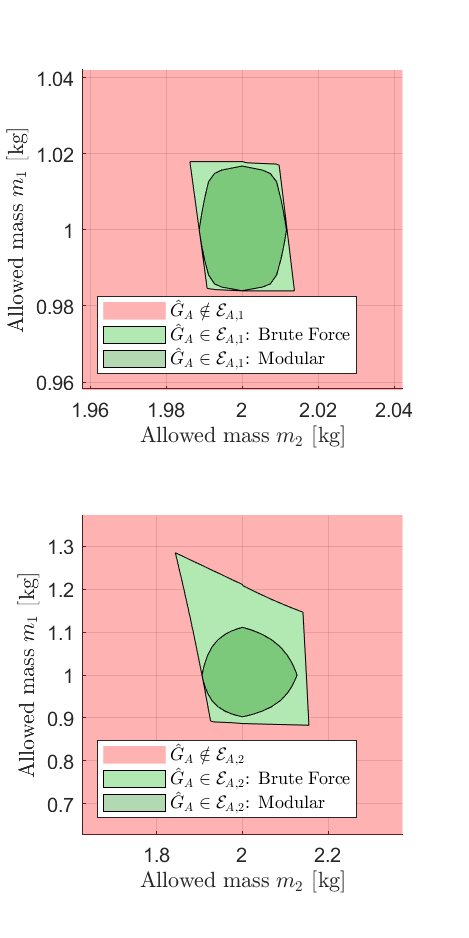}
    \caption{Allowed masses of modules using the brute-force and modular approaches for relatively more $\E_{A,1}$ (top, $\gamma(\omega) = 0.05$ for all $\omega \in \Omega$) and less $\E_{A,2}$ (bottom, $\gamma(\omega) = 0.5$ for all $\omega \in \Omega$) restrictive system FRF specifications.}
    \label{fig:2mass_gammas}
\end{figure}
We compare in Figure~\ref{fig:2mass_gammas} the results for two different system-level FRF specifications, i.e., $\gamma(\omega) = 0.05$ $(\E_{A,1})$ and $\gamma(\omega) = 0.5$ $(\E_{A,2})$ for all $\omega \in \Omega$, respectively.

From Figure~\ref{fig:2mass_gammas}, we can observe two trends that influence the conservativeness of the modular approach.
First, in this example, it can be seen in the brute force approach that (to some minor extent) reducing $m_1$ allows for a further increase in $m_2$ and vice-versa. 
However, in the modular approach, by distributing the FRF specifications from system level towards FRF specifications on module level using Theorem~\ref{the:top_down}, this cannot be taken into account.
Second, for a small $\gamma$, i.e., in the top plot in Figure~\ref{fig:2mass_gammas}, the conservatism of the modular approach is limited compared to the brute force approach.
However, when comparing the top and bottom plot in Figure~\ref{fig:2mass_gammas}, it is clear that allowing more change the system FRF, specifically, 50\%, the conservatism of the modular approach also increases.
This can be explained by the fact that if we allow more change in the individual module FRFs, the total potential effect of all these changes could propagate in the total system.
Therefore, in these cases, to be able to guarantee $\hat{G}_A \in \E_A$, more conservatism is required.

\begin{figure}
  	\centering
   	\includegraphics[scale=.7]{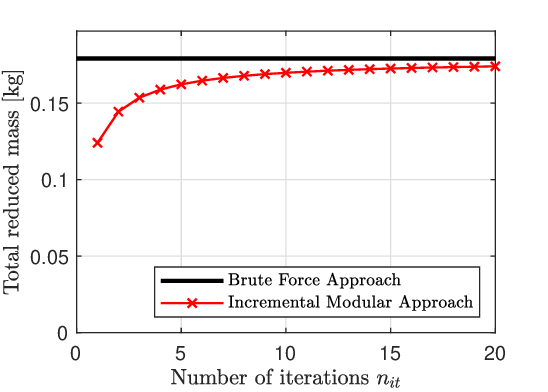}
   	\caption{Total allowed masses reduction satisfying $\hat{G}_A \in \E_A$ where $\gamma(\omega) = 0.5$ for all $\omega \in \Omega$. Comparison between the brute-force approach and the incremental modular approach, where for each iteration, the total mass is reduced given $\gamma(\omega) = 0.5/n_{it}$ for all $\omega \in \Omega$.}
    \label{fig:two_mass_1_it}
\end{figure}

We can mitigate both causes of conservatism using incremental redesign. 
For example, instead of allowing a total change $\gamma(\omega) = 0.5$ for all $\omega \in \Omega$, we could allow a redesign approach in which we allow ten redesign iterations where in each iteration $\gamma(\omega) = 0.05$ for all $\omega \in \Omega$ is used for the system specification $\E_A$.
We show the results of such an incremental approach in Figure~\ref{fig:two_mass_1_it} for various numbers of iterations.
Here, in contrast to the previous results, we aim to reduce the total mass ($m_1+m_2$) in the system as much as possible compare the modular and integrated approach with incremental variations.
From these results, it is clear that using an incremental redesign approach can mitigate most of the conservatism introduced by applying parallel redesign.
In practice, obviously, an incremental approach has the disadvantage that it will increase the total computation time.

Summarizing, the proposed modular approach allows for fully independent redesign of modules.
However, this approach comes at cost of some conservatism. 
Nevertheless, for relatively small changes to the system design, the level of conservativeness is certainly acceptable.
To allow larger overall changes to the system, an incremental redesign approach can be used clearly mitigating the conservatism of the modular redesign approach.

\subsection{Plate Pillar model}
\begin{figure}
  	\centering
   	\includegraphics[scale=.7, page=6]{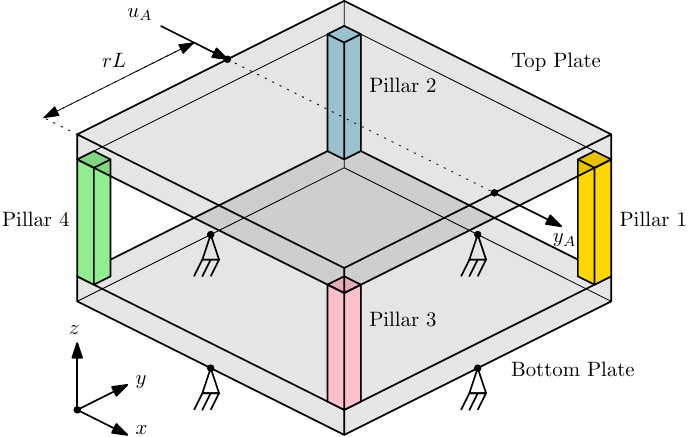}
   	\caption{Schematic drawing of the Matlab-implemented plate pillar model \cite{pillarplatemodel}.}
    \label{fig:plate_pillar_model}
\end{figure}

In this section, we will apply the modular redesign approach to the Matlab-implemented `\verb+platePillarModel+' benchmark system \cite{pillarplatemodel}.
We will use this example system, firstly, to show how the proposed modular approach automatically determines which modules offer more/less freedom for redesign to meet the overall system FRF specifications and, secondly, to illustrate the computational advantage of parallel redesign.
The system is presented in Figure~\ref{fig:plate_pillar_model}.

\begin{figure}
  	\centering
   	\includegraphics[scale=.7]{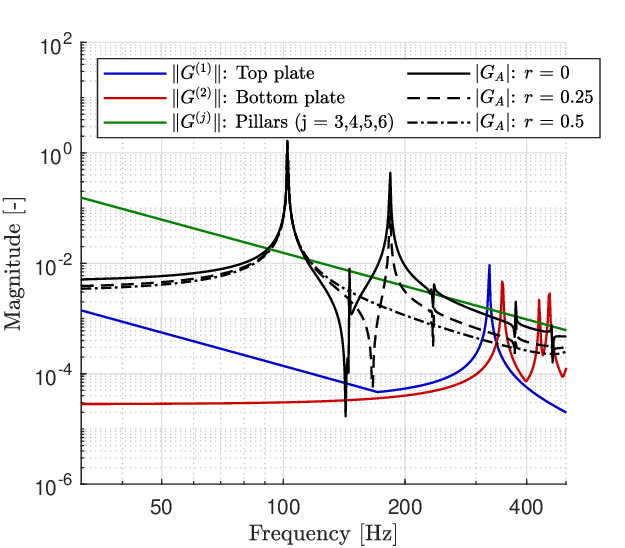}
   	\caption{Matrix 2-norms of the module FRFs $\|G^{(j)}(i\omega)\|$ for $j = 1,\dots,6$ and system FRFs $|G_A(i\omega)|$ for $r = 0$, $r = .25$, and $r = 0.5$.}
    \label{fig:pillar_plate_FRF}
\end{figure}

This system consists of 6 finite-element (FE) modules: four flexible pillars, each consisting of 132 degrees of freedom (DOFs) and 24 interface points, and two flexible plates, each consisting of 2646 DOFs and 48 interface points.\footnote{For the exact system description, see \cite{pillarplatemodel}.}
All pillars are connected at their top to the top plate and at their bottom to the bottom plate.\footnote{The proposed redesign framework uses (a) flexible coupling(s) between modules. Therefore, in this use case, very stiff couplings are used to model almost rigid interconnections between modules, see Remark~\ref{rem:rigid}.}
The bottom plate is connected to the fixed world at the middle of its four bottom edges, see Figure~\ref{fig:plate_pillar_model}.
An external input force $u_A$ is applied in the x-direction to the top edge of the top plate between pillars 2 and 4 at a distance $rL$ ($0\leq r \leq 1, L$ is the distance between pillar 4 and pillar 2) from pillar 4, and an external output displacement $y_A$ is measured in the x-direction at the top edge of the top plate between pillars 1 and 3 at a distance $rL$ from pillar 3. 
The matrix 2-norms of the module FRFs, i.e., $\|G^{(j)}(i\omega)\|$, and the magnitudes of the FRFs of the system $G_A$ from $u_A$ to $y_A$ for $r = 0$, $r = 0.25$, and $r = 0.5$ are given in Figure~\ref{fig:pillar_plate_FRF}.

\begin{figure}
  	\begin{subfigure}{0.5\textwidth}
  		\centering
   	\includegraphics[scale=1,trim={2cm 2cm 2cm 2cm},clip]{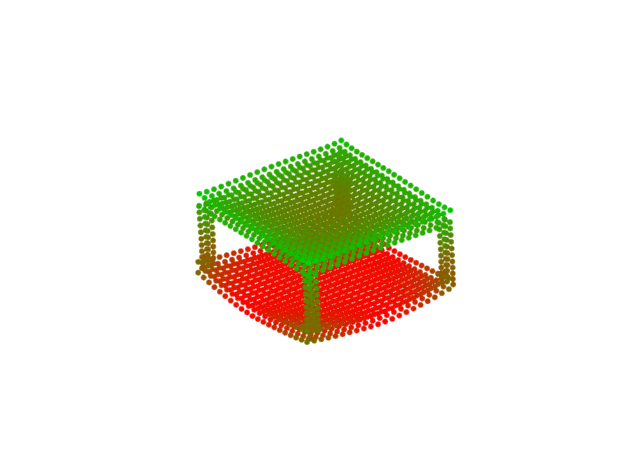}
   		\caption{Operating deflection shape, 5 Hz}
    	\label{fig:op_mode1}
    \end{subfigure}
  	\begin{subfigure}{0.5\textwidth}
  		\centering
   	\includegraphics[scale=1,trim={2cm 2cm 2cm 2cm},clip]{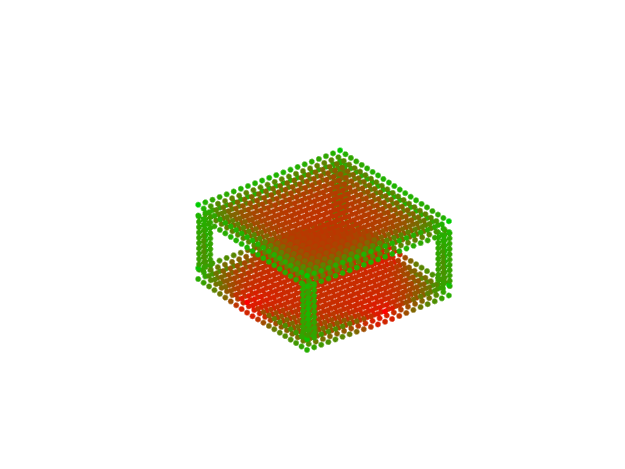}
   		\caption{Operating deflection shape, 120 Hz}
    	\label{fig:op_mode2}
    \end{subfigure}
   	\caption{Operating deflection shapes of the pillar plate model at 5 Hz (a) and 120 Hz (b). Node displacement is visualized with colors ranging from red (low displacement) to green (high displacement).}
    \label{fig:op_modes}
\end{figure}

By definition, as can be seen in Figure~\ref{fig:pillar_plate_FRF}, the pillar modules are identical. 
However, obviously, the effect of each of these pillar modules on the system as a whole does not have to be identical.
To illustrate this, we will investigate the allowed stiffness reduction of the pillars for different frequencies and for different values of $r$.
We will show that following the proposed work-flow in Section~\ref{sec:module_specs} automatically enables us to determine the relative importance of modules with respect to the system FRF specifications.
More specifically, the work-flow for this use case is as follows:
\begin{enumerate}
\item \textbf{Define system specifications}: We define a first system FRF specification $\E_{A,1}(2\pi 5)$ using (\ref{eq:ex_req}) with $\gamma(2\pi 5) = 0.05$, i.e., at the single frequency $\Omega/(2\pi) = 5$ Hz and a second system specification $\E_{A,2}(2\pi 140)$ using (\ref{eq:ex_req}) with $\gamma(2\pi 140) = 0.05$, i.e., at the single frequency $\Omega/(2\pi) = 140$ Hz, both for $r$ ranging from 0 to 1. The operating deflection shapes \cite{richardson1997mode} of the system at these frequencies are illustrated in Figure~\ref{fig:op_modes}.
\item \textbf{Compute module specifications}: Given $\E_{A,1}(2\pi 5)$ and $\E_{A,2}(2\pi 140)$, Algorithm~\ref{alg:top_down} can be used to automatically generate two different module specifications $\E^{(j)}_1(2\pi 5)$ and $\E^{(j)}_2(2\pi 140)$ as in Definition~\ref{def:module_spec} \emph{only} for the pillars, i.e., $j=3,4,5,6$, and not the plates. Practically, this means that we set $W^{(j)}$ and $V^{(j)}$ fixed for $j=1,2$.
\item \textbf{Redesign modules}: Given  $\E^{(j)}_1(2\pi 5)$ and $\E^{(j)}_2(2\pi 140)$ for $j=3,4,5,6$, we investigate how much the stiffness of every pillar could potentially be reduced. Note that in this example, for the sake of simplicity, we reduce the system stiffness by reducing the Young's modules by a percentage. However, replacing the pillar models with a completely new redesigned model $\hat{G}^{(j)}$or $j=3,4,5,6$ is obviously also possible within this framework: the redesigned pillar(s) only need to satisfy the local module FRF specification.
\item \textbf{Integrate redesigned system}: Following from Theorem~\ref{the:top_down}, satisfaction of the overall system FRF specification is guaranteed.
\end{enumerate}

\begin{figure}
  	\centering
   	\includegraphics[scale=.7]{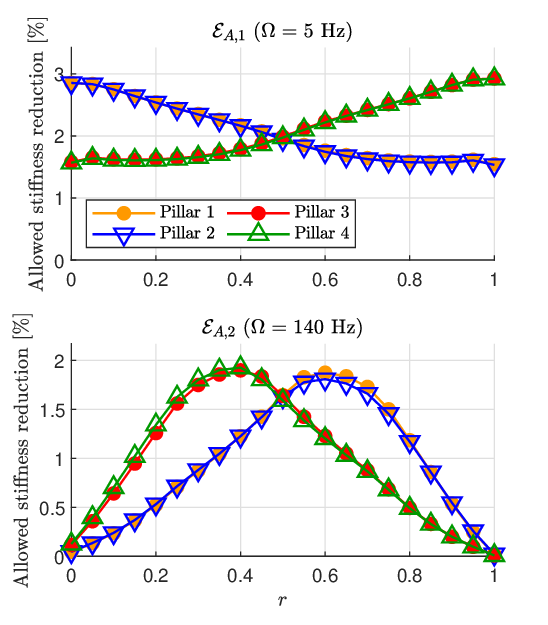}
   	\caption{Allowed stiffness reduction for each pillar based on specifications $\E_{A,1}(2\pi 5)$ (top) and $\E_{A,2}(2\pi 140)$ (bottom) as a function of $r$ (for both, $\gamma(2\pi 5) = 0.05$ and $\gamma(2\pi 140) = 0.05$, respectively, in (\ref{eq:ex_req})).}
    \label{fig:pillar_plate_stiff}
\end{figure}

The results of the two redesigns are given in Figure~\ref{fig:pillar_plate_stiff}.

Figure~\ref{fig:pillar_plate_stiff} shows that, for example, for $\E_{A,1}$, a larger decrease in the stiffness of pillars 1 and 2 is allowed to satisfy the specification at $r=0.2$ compared to pillars 3 and 4 at $r=0.2$, whereas at $\E_{A,2}$ the opposite holds.
This shows that, the proposed redesign method can be used to identify which modules have more/less redesign space to meet the \emph{specific} system FRF specifications, i.e., the sensitivity of module changes with respect to how the overall system FRF specifications are determined.

Additionally, because the redesign can be done independently for each module, computational advantages arise.
Namely, to validate the satisfaction of FRF specifications on a system level, i.e., $\hat{G}_A \in \E_{A}$, $\hat{G}_A$ needs to be computed, which requires interconnection of all redesigned module models ($j=3,4,5,6$) and not redesigned ($j=1,2$) module models $\hat{G}^{(j)}$, and FRF analysis of the redesigned system model.
Instead, when modular redesign is used, only the module FRF specifications $\hat{G}^{(j)} \in \E^{(j)}$ need to be checked, which is computationally significantly less costly.
To show this effect, the FRF computation times of the FRFs of the redesigned modules along with the number of DOFs per module are presented in Table~\ref{tab:comp_times}.\footnote{The computations in this work are applied using on a modern notebook (Intel i7 2.6Ghz Processor, 16Gb RAM).}
\begin{table}[]
\caption{Computation times of the FRF at 100 frequency points of all the (sparse descriptions of the) modules and the overall systems. $^*$Note that computing the FRFs for the top and bottom plate is, as they are not redesigned, not required in the modular redesign approach.}
\label{tab:comp_times}
\begin{tabular}{l|ll}
\textbf{Module} & \textbf{\# of DOFs} & \textbf{FRF calc. time {[}s{]}} \\ \hline
$G^{(1)}$ (Top Plate) & 2646 & 26.65$^*$ \\
$G^{(2)}$ (Bottom Plate) & 2646 & 25.82$^*$ \\
$G^{(3)}$ (Pillar 1) & 132 & 0.90 \\
$G^{(4)}$ (Pillar 2) & 132 & 0.85 \\
$G^{(5)}$ (Pillar 3) & 132 & 0.97 \\
$G^{(6)}$ (Pillar 4) & 132 & 1.03 \\ \hline
$G_A$ & 5820 & 56.22
\end{tabular}
\end{table}

From this table, is is clear that FRF specification validation on a module level is not only more practical from an organizational point of view, it is also significantly cheaper in terms of computational cost.
Especially, when redesign work is done on smaller/less complex modules, such as is this case the pillar modules, the computational gain is significant.
As can be seen in Table~\ref{tab:comp_times}, if a redesigned pillar module needs to be validated on system level, this would take almost a minute, while validating a redesign on module level only requires about a second.
Obviously, the computational advantage becomes larger for larger assembly models.

\subsection{Wire bonder model}
In the final use case, we will show that the proposed redesign approach is applicable to models of industrial-scale mechatronic wire bonder systems.

\begin{figure}
  	\begin{subfigure}{0.5\textwidth}
  		\centering
   		\includegraphics[scale=.2]{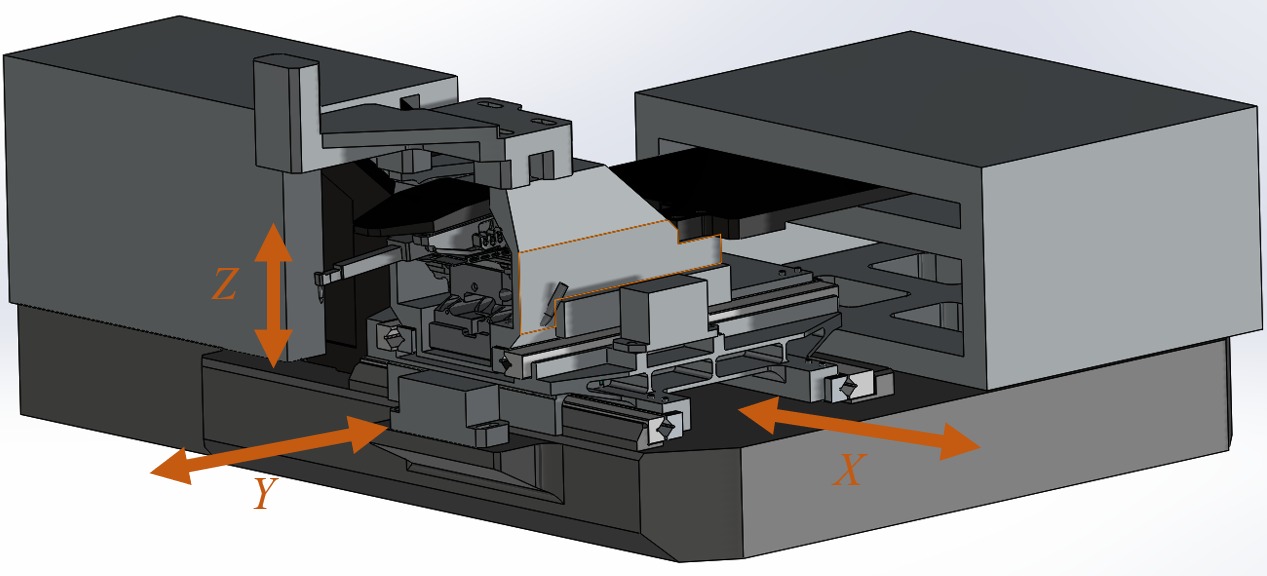}
   		\caption{CAD-model}
    	\label{fig:wirebondera}
    \end{subfigure}
  	\begin{subfigure}{0.5\textwidth}
  		\centering
   		\includegraphics[scale=1, page=8]{figures/IPE_figures.pdf}
   		\caption{Schematic representation}
    	\label{fig:wirebonderb}
    \end{subfigure}
   	\caption{CAD-model (a) and schematic representations (b) of the industrial wire bonder system.}
    \label{fig:wirebonder}
\end{figure}

A wire bonder is a high-performance motion system, used in the semiconductor industry to make electrical connections between the IC chip and the package or substrate that it is mounted on.
To do so, it moves a capillary tip in the x-, y- and z- directions with high precision. 
Figure~\ref{fig:wirebonder} shows the CAD model and a schematic representation. 
The X-stage of the system can move in the x-direction with linear roller slides connected to the machine frame. Furthermore, the Y-stage can move in the y-direction with linear roller slides connected to the X-stage and the z-direction of the capillary tip is positioned with the Z-stage that can rotate through a leaf spring cross-hinge connected to the Y-stage.  

\begin{figure}
  	\centering
   	\includegraphics[scale=.7]{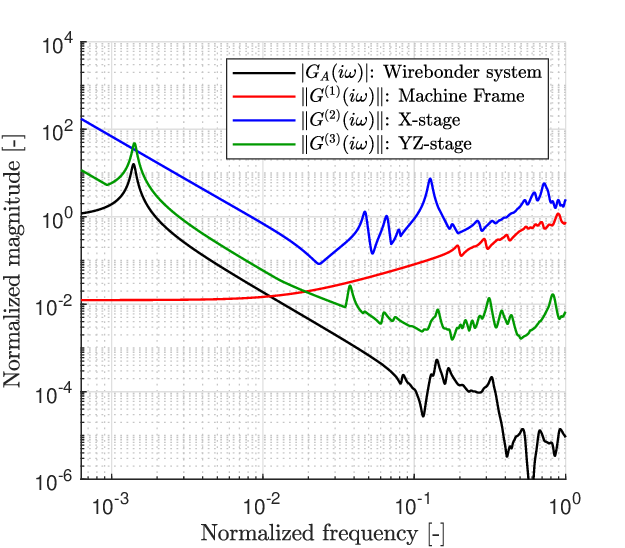}
   	\caption{Magnitude of the wire bonder FRF $|G_A(i\omega)|$ and 2-norm of the module FRF matrices $\|G^{(j)}(i\omega)\|$ for $j = 1,2,3$. For confidentiality, the magnitude and frequency axes are normalized.}
    \label{fig:wirebonder_FRF}
\end{figure}

For the modular model of this system, all modules of this system are modelled as MIMO flexible FE-models with interface DOFs at the interconnection DOFs and at the external input and output DOFs.
Specifically, we define the following three modules:
\begin{enumerate}
\item \textbf{Machine Frame:} The machine frame model is connected to the fixed world at several locations at the bottom of the model. Furthermore, the interface, i.e., the rollers in the two linear roller slides (18 per slide) are modelled as spring-damper interconnections to the X-stage.
\item \textbf{X-stage:} The X-stage model is connected to the machine frame  (described already at 1.) and to the YZ-stage model. 
The interface to the YZ-stage also consists of two linear roller slides (again 18 rollers per slide). 
Again, each roller is modelled as a spring-damper interconnection.
\item \textbf{YZ-stage:} The YZ-stage model is the combination of the Y-stage and the Z-stage. 
An external input force $u_A$ [N] in the z-direction is realized by an electro motor and the resulting vertical output displacement $y_A$ [m] is defined at the end of the capillary tip.
\end{enumerate}
The FRF of the wire bonder system from $u_A$ to $y_A$ is given by $G_A(i\omega)$ and the (2-norms of the FRF matrices of the MIMO) modules are given by $\|G^{(j)}(i\omega)\|$ for $j=1,2,3$ for the machine frame, X-stage, and YZ-stage, respectively.
These quantities are shown in Figure~\ref{fig:wirebonder_FRF}.
\begin{remark}
In practice, the wire bonder model has 6 external inputs: 3 motor forces in the three (translational x, translational y and rotational x) controller directions and 3 reaction forces, and at least 6 external outputs: 3 measured signals at the encoder locations and the x-, y-, and z-position of the capillary tip (in the model, as they cannot be easily measured in the physical system). This full 6-by-6 matrix FRF could be used as the model on which we enforce specifications within the proposed redesign framework.
However, to illustrate the framework clearly and concisely, we focus only on the SISO FRF from the z-input force to the z-position of the capillary tip as in $G_A(i\omega)$.
\end{remark}

\begin{figure}
  	\centering
   	\includegraphics[scale=.7]{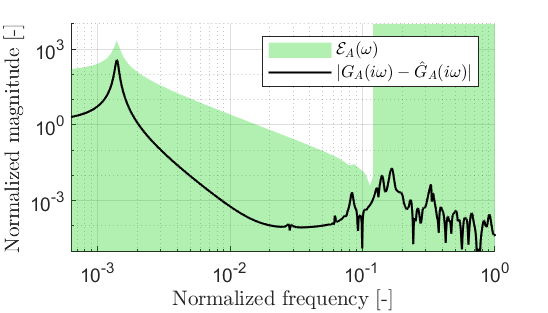}
   	\caption{Overall wire bonder system FRF specification satisfaction $\hat{G}_A \in \E_A$ given the redesigned modules. $|\hat{G}_A(\omega) - {G}_A(\omega)| \in \E_A(\omega)$ for all $\omega$ indicates that the specification is satisfied.}
    \label{fig:wirebonder_result}
\end{figure}

Also in this example, we will apply the work-flow proposed in Section~\ref{sec:module_specs}:
\begin{enumerate}
\item \textbf{Define system specifications}: As the system FRF specification $\E_A(\omega)$, we again use a bound on the maximum allowed relative change as defined in (\ref{eq:ex_req}). 
Here, we select $\gamma(\omega) = 0.1$ for all $\omega\in\Omega$ and we consider the set $\Omega$ to be 100 logarithmically spaced frequency-points from $5\times10^-4$ up to a cut-off frequency of $0.12$ [normalized frequency], which is visualized by the green area in Figure~\ref{fig:wirebonder_result}.
\item \textbf{Compute module specifications}: Algorithm~\ref{alg:top_down} is applied to compute $\E^{(j)}(\omega)$ for $j=1,2,3$ for all $\omega \in \Omega$.
\item \textbf{Redesign modules}: Given $\E^{(j)}$ for all modules, we will propose six redesigns $\hat{G}^{(j)}_i$ for $i=1,\dots,6$ and validate locally if $\hat{G}^{(j)}_i \in \E^{(j)}$ is satisfied. 
The proposed design changes and whether they satisfy their respective FRF specifications are given in Table~\ref{tab:wirebonder}.
\item \textbf{Integrate system}: Following from Theorem~\ref{the:top_down}, if all the selected module redesigns FRFs $\hat{G}^{(j)}_i$ satisfy $\hat{G}^{(j)}_i \in \E^{(j)}$,
satisfaction of the overall system specification $\E_A(\omega)$ is guaranteed. For example, if a lower quality steel with 1\% less stiffness is used in the machine frame, a 15~g encoder is added to the middle of the X-stage and a 5~g sensor is added to the cantilever tip (implemented simultaneously), we can still guarantee that the system specifications are satisfied. This is shown in Figure~\ref{fig:wirebonder_result}.
\end{enumerate}
\begin{table}[]
\caption{Some proposed wirebonder module redesign changes, the figures illustrate the proposed design change at each individual module. If $\hat{G}^{(j)}_i \in \E^{(j)}$, the obtained module specification is satisfied.}
\label{tab:wirebonder}
\begin{tabular}[t]{|c|ll|}
\hline
\rowcolor[HTML]{ffc0cb} 
\textbf{Machine Frame:} &  Change & $\hat{G}^{(1)}_i \in \E^{(1)}$\\ \hline
\begin{tabular}[t]{@{}c@{}} Added point mass \\ \includegraphics[scale=.7, page=14]{figures/IPE_figures.pdf} \end{tabular} &
\begin{tabular}[t]{@{}l@{}} 20 g \\ 200 g \\ 2000 g \end{tabular} &
\begin{tabular}[t]{@{}l@{}} Yes \\ Yes \\ Yes \end{tabular} \\ \hline
\begin{tabular}[t]{@{}c@{}} Reduced Young's modules \\ \includegraphics[scale=.7, page=13]{figures/IPE_figures.pdf} \end{tabular} &
\begin{tabular}[t]{@{}l@{}} 1 \% \\ 2 \% \\ 3 \% \\ \end{tabular} &
\begin{tabular}[t]{@{}l@{}} Yes \\ No \\ No \\ \end{tabular} \\ \hline
\rowcolor[HTML]{add8e6} 
\textbf{X-stage:} &  Change & $\hat{G}^{(2)}_i \in \E^{(2)}$\\ \hline
\begin{tabular}[t]{@{}c@{}} Added point mass \\ \includegraphics[scale=.7, page=12]{figures/IPE_figures.pdf} \end{tabular} &
\begin{tabular}[t]{@{}l@{}} 10 g \\ 15 g \\ 20 g \\ \end{tabular} &
\begin{tabular}[t]{@{}l@{}} Yes \\ Yes \\ No \\ \end{tabular} \\ \hline
\begin{tabular}[t]{@{}c@{}} Added point mass \\ \includegraphics[scale=.7, page=11]{figures/IPE_figures.pdf} \end{tabular} &
\begin{tabular}[t]{@{}l@{}} 10 g \\ 15 g \\ 20 g \\ \end{tabular} &
\begin{tabular}[t]{@{}l@{}} Yes \\ No \\ No \\ \end{tabular} \\ \hline
\rowcolor[HTML]{90ee90} 
\textbf{YZ-stage:} &  Change & $\hat{G}^{(3)}_i \in \E^{(3)}$\\ \hline
\begin{tabular}[t]{@{}c@{}} Reduced stiffness \\ \includegraphics[scale=.7, page=10]{figures/IPE_figures.pdf} \end{tabular} &
\begin{tabular}[t]{@{}l@{}} 2 \% \\ 1 \% \\ 0.1 \% \end{tabular} &
\begin{tabular}[t]{@{}l@{}} No \\ No \\ No \end{tabular} \\ \hline
\begin{tabular}[t]{@{}c@{}} Added point mass \\ \includegraphics[scale=.7, page=9]{figures/IPE_figures.pdf} \end{tabular} &
\begin{tabular}[t]{@{}l@{}} 5 g \\ 10 g \\ 20 g \\ \end{tabular} &
\begin{tabular}[t]{@{}l@{}} Yes \\ No \\ No \\\end{tabular} \\ \hline
\end{tabular}
\end{table}
This example shows that the proposed modular redesign framework can be applied to industrial-scale models.

By defining the system FRF specification as in (\ref{eq:ex_req}) with $\gamma(\omega) = 0.1$ for all $\omega\in\Omega$, we practically allow a maximum change in the FRF at \emph{all} of the frequency point to be 10 \% of the original system FRF $G_A(i\omega)$.
Changing some components of the system will quickly exceed this specification for some frequencies, as a small change in eigenfrequency can already invalidate the specification.
A solution could be to increase $\gamma(\omega)$ at frequencies that are less relevant to the system requirements.

\begin{figure}
  	\centering
   	\includegraphics[scale=.7]{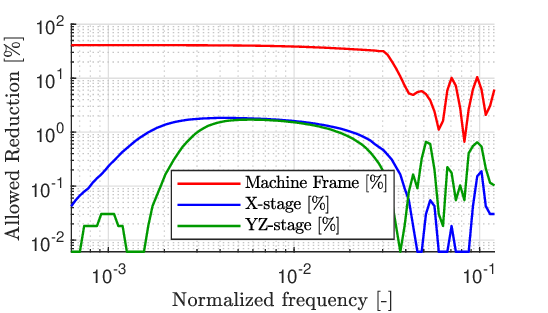}
   	\caption{Allowed Young's Modulus reduction for each of the modules given the satisfaction of specification $\E^{(j)}(\omega)$ \emph{only} at a single $\omega$ instead of at all $\omega \in \Omega$.}
    \label{fig:wirebonder_freqs}
\end{figure}

To analyse the frequency-dependency of the system, in Figure~\ref{fig:wirebonder_freqs}, we show what happens if we only select a single frequency point $\omega$ as specification (with $\gamma(\omega) = 0.1$).
From this figure it can be seen that, at some frequencies, only very small changes to the dynamics are allowed for the X-stage and the YZ-stage.
In addition, although the machine frame can be modified much more, the allowed change in stiffness of the machine frame is still very sensitive to the choice of this frequency point $\omega$.
Therefore, it is important that in the definition of the frequency-dependent specifications, it is taken into account which frequencies are actually essential for the performance of the system and only define requirements at these frequencies.
\begin{remark}
Note that the computational cost of solving the optimization in Algorithm~\ref{alg:top_down}, for the three case studies investigated in this section, is as follows (notebook with Intel i7 2.6Ghz Processor, 16Gb RAM): The optimization, per frequency point, of Algorithm~\ref{alg:top_down} takes circa 0.1, 3 and 40 seconds for the two DOF mass-spring-damper system, plate pillar model and wirebonder system, respectively.
\end{remark}
Summarizing, we have illustrated using three use cases that the proposed modular design framework can be applied for many different applications.
The examples show, given the user-defined system specification for redesign, that the level of conservativeness is typically acceptable, and, if not, an incremental approach can be used.
Furthermore, we have shown that the approach can be directly applied to the complex models industrial-scale systems.

\section{Conclusions}
\label{sec:conclusion}
In this paper, we have presented a complete framework to achieve a fully modular redesign approach with system-level, FRF-based specification guarantees on the dynamics of complex mechatronic systems.
This framework allows for engineering teams to work independently and in parallel and streamlines the replacement of system components.

To enable this modular redesign approach, we have introduced a modular modeling approach using coupled FRFs of individual modules.
Furthermore, we have demonstrated how module FRF specifications can be computed to ensure the satisfaction of overall system FRF specifications.
Both the system and module specifications are defined in terms of a flexible, user-defined, bound on the allowed changes in the FRF of the redesigned models with respect to the original models.
This modular approach enables the independent analysis and verification of proposed changes in individual modules, allowing design teams to work concurrently and efficiently.
We provide the necessary technical results and algorithms and present a work-flow that facilitates the modular redesign framework.

To validate this framework, we have shown practical demonstrations on three (mechatronic) use cases, including a two-mass-spring-damper system, a pillar-and-plate benchmark model, and an industrial wire bonder system. 
These examples showcase the applicability and flexibility of the proposed approach in mechatronic system design and redesign.

\section{Acknowledgments}
\noindent This publication is part of the project Digital Twin with project number P18-03 of the research programme Perspectief which is (mainly) financed by the Dutch Research Council (NWO).

\bibliography{Redesign}  

\begin{thebibliography}{10}

\bibitem{nielsen2015systems}
Claus~Ballegaard Nielsen, Peter~Gorm Larsen, John Fitzgerald, Jim Woodcock, and
  Jan Peleska.
\newblock Systems of systems engineering: basic concepts, model-based
  techniques, and research directions.
\newblock {\em ACM Computing Surveys (CSUR)}, 48(2):1--41, 2015.

\bibitem{baldwin2000design}
Carliss~Young Baldwin and Kim~B Clark.
\newblock {\em Design rules: The power of modularity}, volume~1.
\newblock MIT press, 2000.

\bibitem{zheng2017multidisciplinary}
Chen Zheng, Peter Hehenberger, Julien Le~Duigou, Matthieu Bricogne, and
  Beno{\^\i}t Eynard.
\newblock Multidisciplinary design methodology for mechatronic systems based on
  interface model.
\newblock {\em Research in Engineering Design}, 28:333--356, 2017.

\bibitem{van2010modular}
Thom~J Van~Beek, Mustafa~S Erden, and Tetsuo Tomiyama.
\newblock Modular design of mechatronic systems with function modeling.
\newblock {\em Mechatronics}, 20(8):850--863, 2010.

\bibitem{hamraz2012multidomain}
Bahram Hamraz, Nicholas H.~M. Caldwell, and P.~John~Clarkson.
\newblock {A Multidomain Engineering Change Propagation Model to Support
  Uncertainty Reduction and Risk Management in Design}.
\newblock {\em Journal of Mechanical Design}, 134(10):100905, 09 2012.

\bibitem{giese2004modular}
Holger Giese, Sven Burmester, Wilhelm Sch{\"a}fer, and Oliver Oberschelp.
\newblock Modular design and verification of component-based mechatronic
  systems with online-reconfiguration.
\newblock In {\em Proceedings of the 12th ACM SIGSOFT twelfth international
  symposium on Foundations of software engineering}, pages 179--188, 2004.

\bibitem{habib2014comparative}
Tufail Habib and Hitoshi Komoto.
\newblock Comparative analysis of design concepts of mechatronics systems with
  a {CAD} tool for system architecting.
\newblock {\em Mechatronics}, 24(7):788--804, 2014.

\bibitem{barbieri2014model}
Giacomo Barbieri, Cesare Fantuzzi, and Roberto Borsari.
\newblock A model-based design methodology for the development of mechatronic
  systems.
\newblock {\em Mechatronics}, 24(7):833--843, 2014.

\bibitem{zheng2019interface}
Chen Zheng, Xiansheng Qin, Beno{\^\i}t Eynard, Jing Li, Jing Bai, Yicha Zhang,
  and Samuel Gomes.
\newblock Interface model-based configuration design of mechatronic systems for
  industrial manufacturing applications.
\newblock {\em Robotics and Computer-Integrated Manufacturing}, 59:373--384,
  2019.

\bibitem{grimheden2013can}
Martin~Edin Grimheden.
\newblock Can agile methods enhance mechatronics design education?
\newblock {\em Mechatronics}, 23(8):967--973, 2013.

\bibitem{de2021process}
Rogerio~Atem de~Carvalho, Henrique da~Hora, and Rodrigo Fernandes.
\newblock A process for designing innovative mechatronic products.
\newblock {\em International Journal of Production Economics}, 231:107887,
  2021.

\bibitem{maier1998architecting}
Mark~W Maier.
\newblock Architecting principles for systems-of-systems.
\newblock {\em Systems Engineering: The Journal of the International Council on
  Systems Engineering}, 1(4):267--284, 1998.

\bibitem{janssen2023modular}
Lars A~L Janssen, Bart Besselink, Rob H~B Fey, and Nathan {van de Wouw}.
\newblock Modular model reduction of interconnected systems: A top-down
  approach.
\newblock {\em IFAC-PapersOnLine}, 56(2):4246--4251, 2023.
\newblock 22nd IFAC World Congress.

\bibitem{janssen2022modular}
Lars A~L Janssen, Bart Besselink, Rob H~B Fey, and Nathan van~de Wouw.
\newblock Modular model reduction of interconnected systems: A robust
  performance analysis perspective.
\newblock {\em Automatica}, 160:111423, 2024.

\bibitem{janssen2022modeselect}
Lars~AL Janssen, Rob~HB Fey, Bart Besselink, and Nathan van~de Wouw.
\newblock Mode selection for component mode synthesis with guaranteed assembly
  accuracy.
\newblock {\em arXiv preprint arXiv:2310.17320}, 2023.

\bibitem{iwasaki2005time}
Tetsuya Iwasaki, Shinji Hara, and Alexander~L Fradkov.
\newblock Time domain interpretations of frequency domain inequalities on
  (semi) finite ranges.
\newblock {\em Systems \& control letters}, 54(7):681--691, 2005.

\bibitem{iwasaki2007feedback}
T~Iwasaki and S~Hara.
\newblock Feedback control synthesis of multiple frequency domain
  specifications via generalized kyp lemma.
\newblock {\em International Journal of Robust and Nonlinear Control:
  IFAC-Affiliated Journal}, 17(5-6):415--434, 2007.

\bibitem{nordebo1999semi}
Zhuquan Zang.
\newblock Semi-infinite linear programming: A unified approach to digital
  filter design with time-and frequency-domain specifications.
\newblock {\em IEEE Transactions on Circuits and Systems II: Analog and Digital
  Signal Processing}, 46(6):765--775, 1999.

\bibitem{zhou1998essentials}
Kemin Zhou and John~Comstock Doyle.
\newblock {\em Essentials of robust control}, volume 104.
\newblock Prentice hall Upper Saddle River, NJ, 1998.

\bibitem{sandberg2009model}
Henrik Sandberg and Richard~M Murray.
\newblock Model reduction of interconnected linear systems.
\newblock {\em Optimal Control Applications and Methods}, 30(3):225--245, 2009.

\bibitem{reis2008survey}
Timo Reis and Tatjana Stykel.
\newblock A survey on model reduction of coupled systems.
\newblock In {\em Model order reduction: theory, research aspects and
  applications}, pages 133--155. Springer, 2008.

\bibitem{craig2000coupling}
Roy~R Craig~Jr.
\newblock Coupling of substructures for dynamic analyses-an overview.
\newblock In {\em Proceedings of the 41st Structures, Structural dynamics, and
  materials conference and exhibit}, page 1573, 2000.

\bibitem{de2008general}
Dennis De~Klerk, Daniel~J Rixen, and S~N Voormeeren.
\newblock General framework for dynamic substructuring: history, review and
  classification of techniques.
\newblock {\em AIAA journal}, 46(5):1169--1181, 2008.

\bibitem{packard1993complex}
Andrew Packard and John Doyle.
\newblock The complex structured singular value.
\newblock {\em Automatica}, 29(1):71--109, 1993.

\bibitem{zhang2006schur}
Fuzhen Zhang.
\newblock {\em The Schur complement and its applications}, volume~4.
\newblock Springer Science \& Business Media, 2006.

\bibitem{pillarplatemodel}
interface: Specify physical connections between components of mechss model.
\newblock \url{https://mathworks.com/help/control/ref/mechss.interface.html}.
\newblock Accessed: 2023-12-01.

\bibitem{richardson1997mode}
Mark~H Richardson et~al.
\newblock Is it a mode shape, or an operating deflection shape?
\newblock {\em Sound and Vibration}, 31(1):54--67, 1997.

\end{thebibliography}

\appendix
\section{System specification examples}
\label{app:ex}
In this appendix, some additional clarifications and several practical examples of system FRF specification are given to show how the required weighting matrices $V_A(\omega)$ and $W_A(\omega)$ are obtained for the modular redesign approach.
As given in (\ref{eq:Ec_bound}) in Definition~\ref{def:system_spec}, the system FRF specifications are satisfied, i.e., $G_A(i\omega)\in\E_A(\omega)$, if
\begin{align}
\label{eq:A_0}
\|V_A(\omega)(G_A(i\omega)-\hat{G}_A(i\omega))W_A(\omega)\| < 1.
\end{align}
where $V_A(\omega)$ and $W_A(\omega)$ are diagonal positive matrices with dimensions $m_A\times m_A$ and $p_A\times p_A$, respectively.
In addition $(G_A(i\omega)-\hat{G}_A(i\omega))$ is a complex matrix of $m_A\times p_A$.

In case of a SISO system, i.e., $m_A = p_A = 1$, (\ref{eq:A_0}) implies
\begin{align}
|G_A(i\omega)-\hat{G}_A(i\omega)| < V_A(\omega)W_A(\omega).
\end{align}
Therefore, in this case, any specification $\gamma$ on the magnitude of $G_A(i\omega)-\hat{G}_A(i\omega)$ can simply be $V_A(\omega) = W_A(\omega) = \sqrt{\gamma}$.

Now consider an example multi-input single-output system, with $m_A = 3$ and $p_A = 1$. 
We want to define an FRF specification in which we want to specify, as an example, that the response to the first input cannot be changed as much due to the redesign as compared to the other two inputs. 
In this case, we could define $V_A(\omega) = \text{diag}\left(3,1,1\right)$ and $W_A = 1$, then, a change to the response to the first input in the FRF $(G_A(i\omega)-\hat{G}_A(i\omega))$ contributes to system FRF specifications three times more than the other 2 inputs.
Automatically, in the redesign of the system, changes to the system that contribute to changes in the response to the first input will then exceed the requirements easier compared to the other two inputs.
For MIMO system in general, the elements in $V_A(\omega)$ and $W_A(\omega)$ can be similarly defined to obtain specifications that allow for differentiation between specific input-output pairs.

Finally, we consider a general MIMO case in which we want to ensure that
\begin{align}
\label{eq:A_1}
\|(G_A(i\omega)-\hat{G}_A(i\omega))\| < \gamma,
\end{align}
independent of the contributions of the individual input-output pair in the system.
In this case, $G_A(i\omega)\in\E_A(\omega)$ as in (\ref{eq:Ec_bound}) guarantees that (\ref{eq:A_1}) is satisfied if
\begin{align*}
V_A(\omega) = \gamma^{-\frac{1}{2}}I, \text{ and } W_A(\omega) = \gamma^{-\frac{1}{2}}I
\end{align*}
of appropriate dimensions.
Note that this is shown trivially by substitution $\gamma^{-\frac{1}{2}}I$ in (\ref{eq:A_0}) to obtain
\begin{align*}
\|\gamma^{-\frac{1}{2}}I(G_A(i\omega)-\hat{G}_A(i\omega))I\gamma^{-\frac{1}{2}}\| &< 1, \\
\frac{1}{\gamma}\|G_A(i\omega)-\hat{G}_A(i\omega)\| &< 1, \\
\|G_A(i\omega)-\hat{G}_A(i\omega)\| &< \gamma.
\end{align*}
This way, any $\gamma$ can be arbitrarily selected to enforce (\ref{eq:A_1}) at any frequency $\omega$ and the specifications in terms of $V_A(\omega)$ and $W_A(\omega)$ can be obtained for the proposed modular redesign approach in this paper.
\end{document}